\documentclass[journal]{IEEEtran}
\usepackage{amsmath,amsfonts}
\usepackage{algorithmic}
\usepackage{algorithm}
\usepackage{array}
\usepackage[caption=false,font=normalsize,labelfont=sf,textfont=sf]{subfig}
\usepackage{textcomp}
\usepackage{stfloats}
\usepackage{url}
\usepackage{verbatim}
\usepackage{graphicx}
\usepackage{cite}
\usepackage{lipsum}
\usepackage{mathtools}
\usepackage{cuted}
\usepackage{xcolor}
\usepackage{amssymb}
\DeclareUnicodeCharacter{2212}{-}
\ifCLASSINFOpdf
\else
\fi
\begin{document}
%
\title{Graphene Metamaterials Based Plasmon-Induced Terahertz Modulator for High-Performance Multiband Filtering and Slow Light Applications}
%
%
%

\author{Dip Sarker$^1$, ~\IEEEmembership{Member,~IEEE},
        Partha Pratim Nakti$^2$, and Ahmed Zubair$^{1, \dagger}$, ~\IEEEmembership{Senior~Member,~IEEE}
\thanks{$^1$~Department
of Electrical and Electronic Engineering, Bangladesh University of Engineering and Technology, Dhaka 1205, Bangladesh.}
\thanks{$^2$~Department
of Electrical and Electronic Engineering, Shahjalal University of Science and Technology, Sylhet 3114, Bangladesh.}
\thanks{$^{\dagger}$~ahmedzubair@eee.buet.ac.bd}}

\maketitle

\begin{abstract}
We proposed multilayered graphene (Gr)-based surface plasmon resonance-induced high-performance terahertz (THz) modulators with tunable resonance frequencies. Several THz plasmonic modulators based on Gr metamaterials were previously reported; however, these modulators had small group delay, low extinction ratio (ER), and difficult-to-tune resonant frequency without structural parameters in the THz range. A comprehensive investigation employing the finite-difference time-domain (FDTD) simulation technique revealed high group delay, broad tunability independent of structural parameters, and large ER for our proposed quadband and pentaband plasmonic modulators. We obtained tunable group delays with a maximum of 1.02 ps and 1.41 ps for our proposed quadband and pentaband plasmonic modulators, respectively, which are substantially greater compared to previously reported Gr-based metamaterial structures. The maximum ER of 22.3 dB was obtained which was substantially high compared to previous reports. Our proposed modulators were sensitive to the polarization angle of incident light; therefore, the transmittance at resonant frequencies was increased while the polarization angle varied from 0$^{\circ}$ to 180$^{\circ}$. These high-performance plasmonic modulators have emerging potential for the design of optical buffers, slow light devices, multistop band filters, integrated photonic circuits, and various optoelectronic systems.  
\end{abstract}
\section{Introduction}
\IEEEPARstart{T}{he} invention of metamaterials has garnered many academic and technological interests due to their extraordinary electromagnetic (EM) properties, which are not easily obtained in the existing photonics and optics world. Plasmonic metamaterials utilize surface plasmons (SPs) which exhibit the most compelling EM properties among the ramifications of the metamaterial~\cite{Pendry}. Structures comprising noble metals, such as gold and silver, can overcome the limitations of classical diffraction limit and realize a photonic loop in the sub-wavelength range of light waves; however, significant energy loss limits practical applicability due to the poor binding effect on incoming light and scattering effect~\cite{Gramotnev2010}. Gr, a two-dimensional material with a hexagonal lattice, has similar properties as noble metals in the THz and infrared regime~\cite{Grigorenko2012}. Most importantly, Gr can excite SPs with strong localization ability, low transmission loss, high tunability, and high group delay. These properties make Gr metamaterial-based structures advantageous in applications including optical cloaking~\cite{D}, perfect lens~\cite{PhysRevLett.85.3966}, modulators~\cite{Sarker2023}, photodetectors~\cite{Liu2014}, polarizers~\cite{Sarker:21}, bandstop filtering~\cite{Li_2022}, and slow light devices~\cite{Xu_2020}.

Bandstop filters are essential components of information processing systems. Recently, plasmonic bandstop filters have received considerable attention from researchers. Conventional metal-based bandstop filters operate well in the visible and near-infrared ranges; however, their applications are limited by the poor SP confinement in the THz frequency range. Contrarily, Gr-based bandstop filters exhibit strong SP confinement providing excellent performance in the THz spectral band. Tunability of resonant frequency without varying structural parameters is another distinguishing attribute for Gr-based structures, whereas tuning characteristics are not easily obtained in metal-based structures where liquid crystals or liquid metals with variable permittivity have been proposed. However, tunning is achievable by simply changing the chemical potential of Gr with an external gate voltage due to Gr's unique energy band structure. This allows fine-tuning of resonant frequencies and group velocities in Gr-based bandstop filters. Graphene nanoribbon (GNR)-Gr disk~\cite{Zhang2013}, multilayer Gr metamaterial~\cite{Zhang:21}, continuously patterned Gr monolayer~\cite{Gao:20, D0CP06182D}, H-shape Gr-metamaterial~\cite{Gao:19}, split-ring resonator~\cite{Wang:22}, and asynchronous crisscross structures~\cite{C8CP04484H} were reported; however, extinction ratio and quality factor, two performance parameters of bandstop plasmonic filter, were quite small. Therefore, there is a scope for better Gr metamaterial-based design to improve the performance of the bandstop optical filters. 

Gr-based plasmonic structures play a vital role in slow light applications, such as optical buffers and quantum memory, by the tuning properties of SPs originated by the interaction of Gr and incoming light. A slow light device can slow down the speed of light pulses propagating through it. The concept behind slow light is based on the phenomenon of light propagation in a medium with a refractive index that can be manipulated by increasing the group delay of light~\cite{doi:10.1080/09500340903159495, Shi:13}. Group delay, a performance parameter of slow light devices, is a measure of how much the light in a medium is delayed compared to the speed of light in the vacuum. Multilayered Gr and different shapes of GNRs exhibited slow light properties due to light-matter interactions; however, the group delay and delay-bandwidth product of the previous reports were low~\cite{Wang:22, Gao:20, Zhang:21}. Zhao \textit{et al.} proposed a ring metamaterial structure comprising noble metals, which provided a high group delay; however, tunability can not be possible without changing structural parameters~\cite{Zhao:19}. Thus, there is a scope for significant improvement in slow light device performance with wide tunability for THz interconnects, quantum memory, and optical buffer applications.  

In this work, we proposed two multiband THz modulator structures comprising GNR, Gr nanoholes, and polymethyl methacrylate (PMMA). We employed the FDTD simulation method to optimize the structures and evaluated the performance parameters of the devices. These structures exhibited dips in their transmittance spectra that arose from the interaction between the incoming light and Gr-based metamaterials. Four- and five-patterned Gr layers on PMMA substrate were utilized for the quadband and pentaband plasmonic modulators, respectively, where each Gr layer acted as an attenuating medium. Our proposed modulators achieved high ER and group delay at resonant frequencies due to strong light-matter interactions making it promising for slow light applications. Moreover, the plasmonic modulators demonstrated high tunability by varying the chemical potentials of Gr layers, which acted as a perfect multi-bandstop filter. We analyzed the polarization sensitivity of our proposed modulators when they were used as bandstop filters. Additionally, we provided a comprehensive comparative study of performance parameters among previously reported Gr-based and our proposed THz modulator structures. 
\section{Methodology}
\begin{figure}[ht] 
\centering
\includegraphics[width=8.87cm]{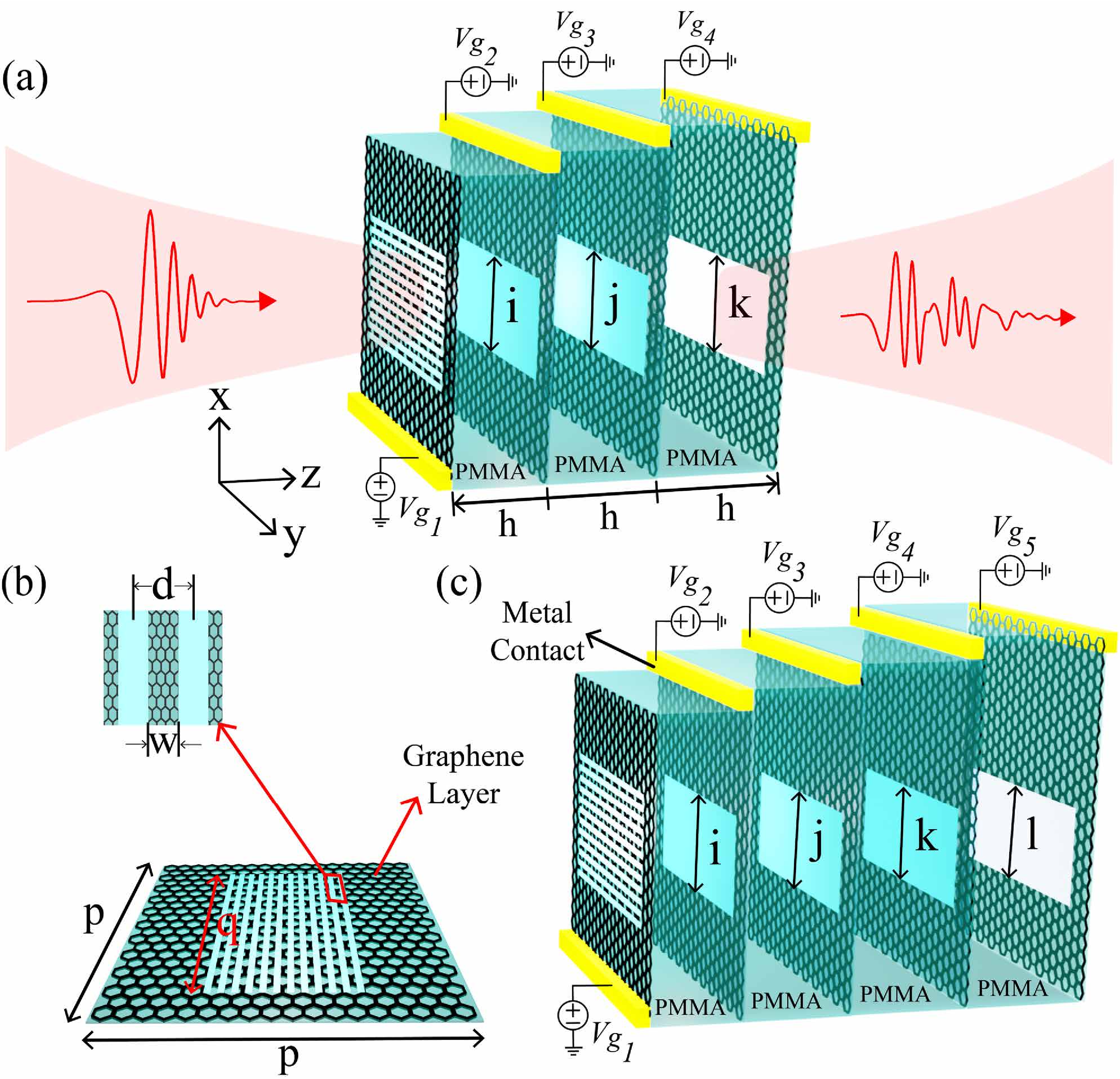}
\caption{(a) Illustration of a quadband plasmonic modulator unit cell. Here, four Gr monolayers were employed, and the structural parameters were: i = j = k = 150 nm, h = 2 $\mu$m. The output transmission signal attenuated and deviated from the source signal due to scattering. (b) Top Gr layer of the proposed structures where the period of the structure, p = 3 $\mu$m, length of GNR, q = 2.5 $\mu$m. Inset shows the magnified view of GNR. Here, the pitch distance of GNR, d is 20 nm, width, w = 10 nm. (c) Schematic view of a pentaband plasmonic modulator unit cell with geometrical parameters: i = j = k = l = 150 nm. $V_{g_n}$ is the applied voltage source of the n-th Gr layer via a metal contact. The $V_{g_n}$ varied the $\mu_{c_n}$ of Gr layers and $\sigma_g$.}
\label{fig:1}
\end{figure}

\begin{figure}[ht] 
\centering\includegraphics[width=8.87cm]{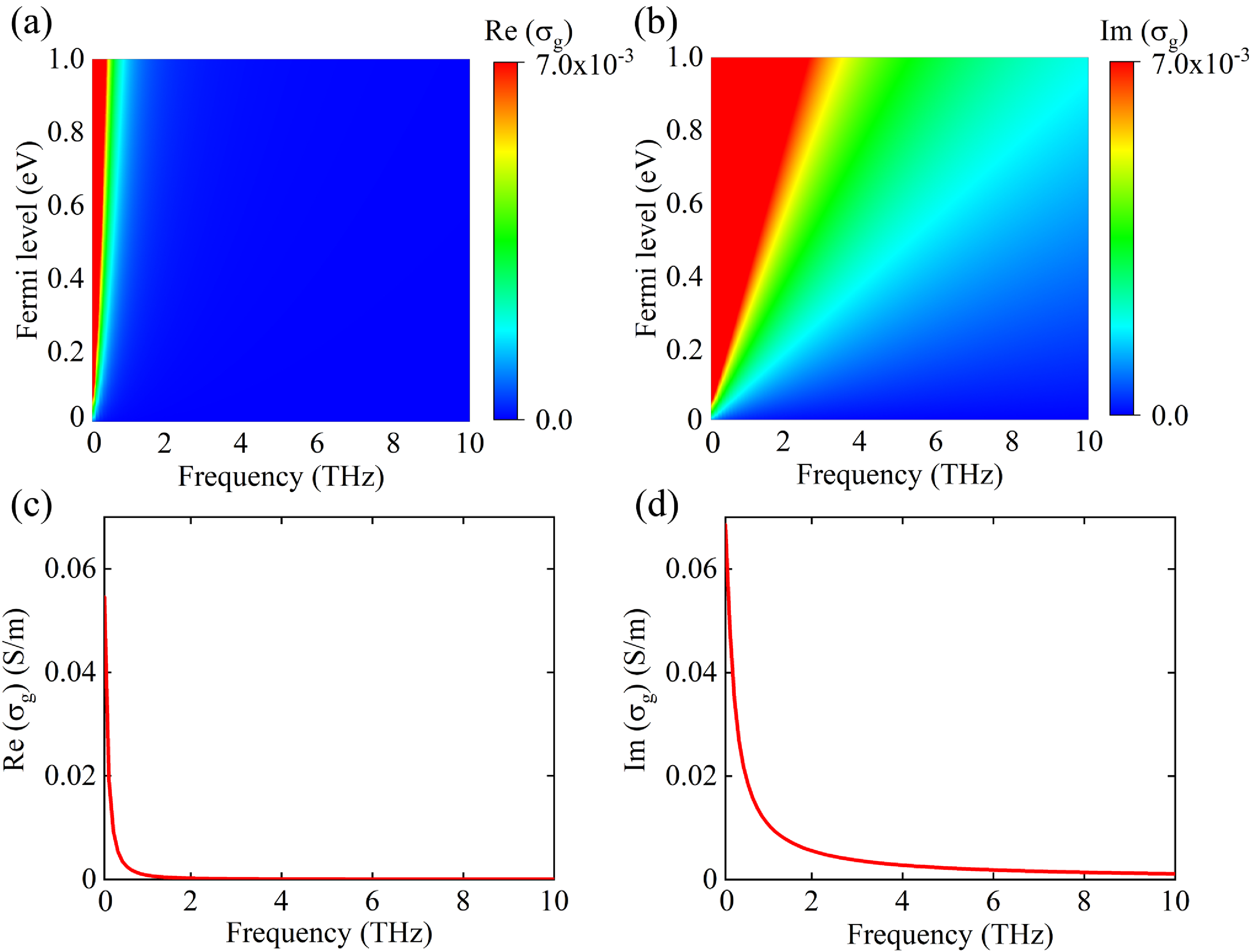}
\caption{Evolution of (a) Re ($\sigma_g$) and (b) Im ($\sigma_g$) of Gr layer at different Fermi levels. (c) Re ($\sigma_g$) and (d) Im ($\sigma_g$) parts of the Gr layer at the $E_F$ of 0.6 eV. Gr exhibited more metallic behavior with increasing Fermi energy levels.}
\label{fig:2}
\end{figure}

A unit cell of the Gr-based plasmonic quadband modulator is illustrated in Fig.~\ref{fig:1}(a). Contrary to the conventional modulator designs, the proposed structure used Gr's excellent SPP and metal-like properties to enhance the modulator's performance. The quadband modulator was designed by comprising four Gr monolayers, which were made up of GNR and Gr nanoholes on PMMA substrate as illustrated in Fig.~\ref{fig:1}(a). Here, the substrate height was selected to be 2 $\mu$m to avoid plasmonic coupling and signal interference. We optimized the GNR arrangement of the modulator's top Gr layer, as depicted in \textcolor{blue}{Fig. S1} of \textcolor{blue}{Supplementary Material}. Twelve periodic airstrips or eleven GNRs on the Gr monolayer were adopted for our proposed modulators. The chemical vapor deposition (CVD) technique can grow a such Gr monolayer on PMMA~\cite{Tai2017}. Fig.~\ref{fig:1}(b) presents the GNR layer of the proposed structures with length, q of 2.5 $\mu$m, width, w of 10 nm, and pitch, d of 20 nm. Inset shows the zoomed view of the GNR layer. The period, p, of the proposed structures, was set to be 3 $\mu$m. The widths of the nanoholes, i = j = k of 150 nm in the rest of the Gr layers, were adopted with length, q of 2.5 $\mu$m. The helium ion beam lithography technique can be utilized to precisely create GNR because of its ability to fabricate GNR below five nm~\cite{Abbas2014}. To design a pentaband plasmonic modulator, we utilized an additional Gr layer with a nanohole below, where the width and length of the Gr nanohole were the same as before. Fig.~\ref{fig:1}(c) illustrates the schematic representation of a unit cell of a pentaband plasmonic modulator. 
The surface current, J, of the Gr layer was modeled by employing the equation of Ohm's law: J = $\sigma_g$E where $\sigma_g$ and E are the Gr's complex surface conductivity and electric field, respectively. We modeled the $\sigma_g$ of Gr by using Kubo's formulation (see \textcolor{blue}{Section 2} of \textcolor{blue}{Supplementary Material})~\cite{Hanson2008}. This formulation was modeled specifically for the $\sigma_g$ of Gr monolayer; however, it can be conducive to explaining multilayer Gr's $\sigma_g$~\cite{Casiraghi2007}. The interband conductivity term can be neglected due to $E_F \gg k_B T$ in the THz frequency regime where $E_F$, $k_B$, and $T$ denote the Fermi energy level, Boltzmann constant, and temperature, respectively. The $\sigma_g$ of Gr is given by,
\begin{equation}
    \sigma_g = \frac{j e^2 E_F}{\pi \hbar^2 (\omega + j \tau^{-1})}.
    \label{Eq.5}
\end{equation}
Here, $e$, $\hbar$, and $\omega$ are the electron charge, reduced Plunck's constant, and angular frequency, respectively. $\tau$ is the relaxation time of the Gr layer, which can be expressed by $\tau = \mu E_F / e{V_{F}}^{2}$. The Fermi velocity, $V_{F}$, and mobility, $\mu$ were set to be 10$^6$ m/s, and 0.4 $m^2V^{−1}s^{−1}$, respectively~\cite{Zhuo:22}. The real and imaginary surface conductivity of Gr as a function of $E_F$ and frequency are shown in Figs.~\ref{fig:2}(a) and (b). The imaginary part of the Gr conductivity rises while the real part stays constant as $E_F$ increases. Therefore, Gr exhibits more metal-like properties with the increase of $E_F$ producing SPPs. Figs.~\ref{fig:2}(c) and (d) illustrate the real and imaginary surface conductivity of the Gr layer at $E_F$ = 0.6 eV. The refractive index data of PMMA was adapted from Lee \textit{et al.}~\cite{Lee2001}.  
We conducted the study utilizing the three-dimensional (3D) FDTD simulation technique. The 3D FDTD simulation range was set to be 3 $\mu$m $\times$ 3 $\mu$m $\times$ 35 $\mu$m. We utilized the periodic boundary conditions in the x- and y-directions, and 24-steep angle perfectly matched layers (PMLs) in the z-direction to absorb light after the simulation region without back reflections. We employed a transverse magnetic (TM) polarized plane-wave light source on the top of the proposed modulators along the z-direction with a center frequency of 3 THz and a frequency span of 5.7 THz. To consider the room temperature for our study, the temperature of the simulation environment was set to 300 K. FDTD is a time-varying and time-consuming technique: larger mesh sizes shorten the computational time which results in low accuracy, and smaller mesh sizes provided high accuracy, increasing computational time and memory size. In our work, smaller mesh sizes were utilized in the region of interest to resolve tiny features, whereas bigger mesh sizes were used in other regions. The non-uniform mesh was utilized to complete simulations in a reasonable amount of time with maintaining the memory requirements. We utilized a mesh size of 8 nm in the GNR and Gr nanoholes for better accuracy in our simulations. Near-field monitors were utilized to obtain the spatial electric field profiles. Transmittance and reflectance spectra were recorded by using power monitors. 
\section{Results and Discussion}
Interaction between the incident plane wave's electric field and free electrons or plasmons of the Gr layer resulted in SP resonance on the surface of the Gr layers of the proposed modulators. As a result, the modulators attenuated light by scattering that created dips in transmittance spectra at resonant frequencies. Based on this phenomenon, we utilized each Gr layer as an attenuating medium at the resonant frequency. Most importantly, this phenomenon will be beneficial for designing tunable plasmonic bandstop filters and slow light devices. 
\subsection{Bandstop filtering applications}
\subsubsection{Quadband filter}
\begin{figure}[H] 
\centering\includegraphics[width=8.87cm]{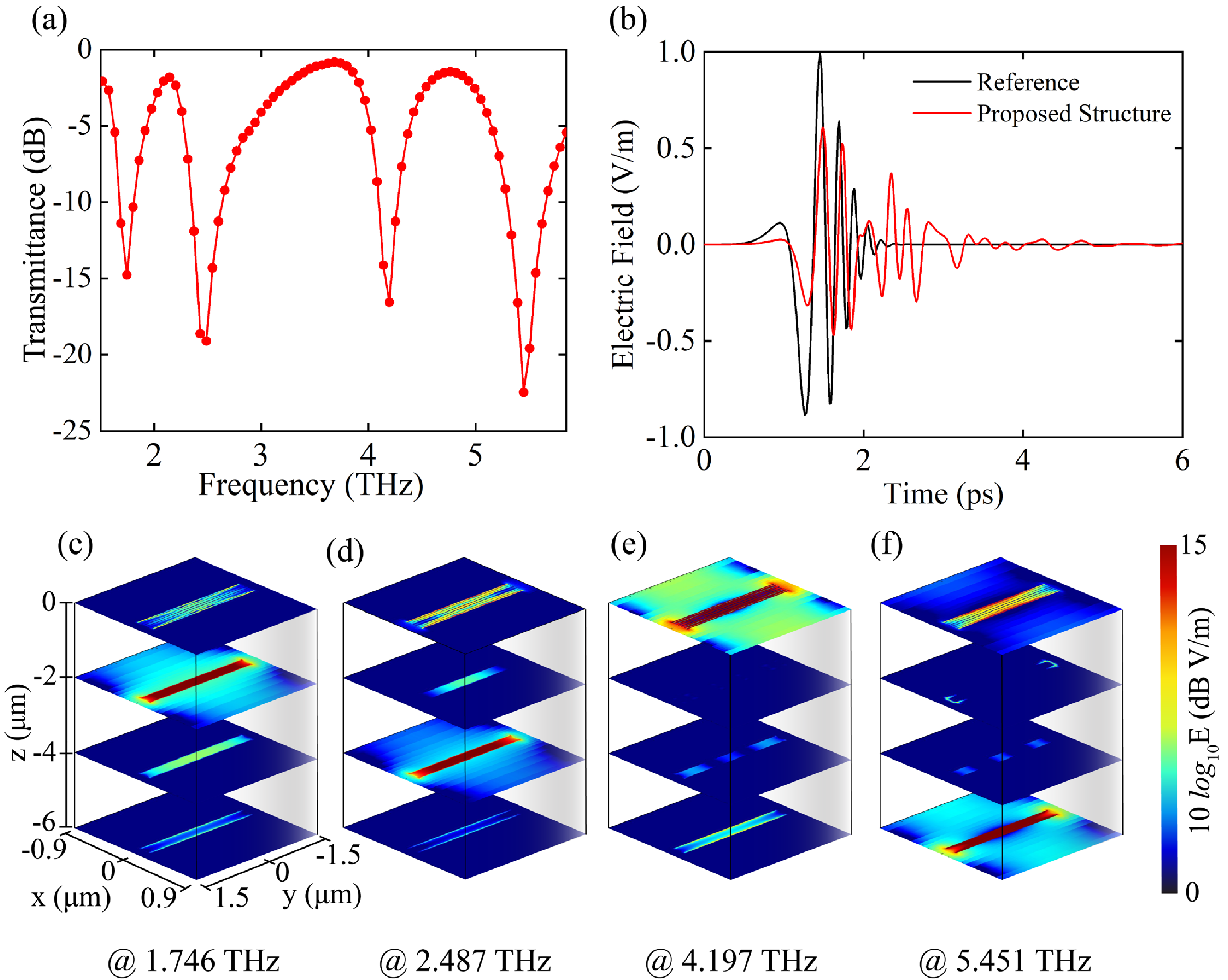}
\caption{(a) Transmittance spectra of the proposed quadband plasmonic filter. Here, $\mu_{c_1}$ = 0.28, $\mu_{c_2}$ = 0.20, $\mu_{c_3}$ = 0.40, and $\mu_{c_4}$ = 0.52 eV were employed. (b) The E-field time-domain waveforms of the proposed quadband plasmonic filter at the normal incidence for TM polarized incident THz light. The transmitted waveform attenuated and deviated from the source THz waveform. The spatial E-field distribution of Gr layers at resonant frequencies of (c) 1.746, (d) 2.487, (e) 4.197, and (f) 5.451 THz for our proposed quadband plasmonic filter. The top, second, third, and bottom Gr layers showed localized E-fields at the resonant frequencies of 4.197, 1.746, 2.487, and 3.385 THz, respectively. The logarithmic scale was utilized to represent the color map of the E-field.}
\label{Quad_filter}
\end{figure}

We utilized the Gr layer's light-blocking phenomenon to obtain the filtering mechanism by enhancing surface light-matter interactions. This blockage of light was the origin of the bandstop plasmonic filter. Based on this mechanism, we utilized four engineered Gr layers with different chemical potentials, providing four transmittance dips due to strong light-matter interaction at resonant frequencies. Fig.~\ref{Quad_filter}(a) depicts the transmittance spectra of the proposed quadband plasmonic filter structure. Here, chemical potentials of Gr layers $\mu_{c_1}$ = 0.28, $\mu_{c_2}$ = 0.20, $\mu_{c_3}$ = 0.40, $\mu_{c_4}$ = 0.52 eV were adopted. Four transmittance dips formed at the resonant frequencies of 1.746, 2.487, 4.197, and 5.451 THz, respectively. Fig.~\ref{Quad_filter}(b) represents the transmitted THz wave in the time domain. The structure experienced a huge deviation from the reference source THz wave because of attenuation by scattering. The transmittance, reflectance, and absorptance spectra are given in \textcolor{blue}{Fig. S2(a)} of \textcolor{blue}{Supplementary Material}. To get a better understanding, we presented the spatial electric field distributions of our proposed plasmonic quadband modulator in Figs.~\ref{Quad_filter}(c)--(f). Each Gr layer primarily localized the electric light field at a resonant frequency. First, second, third, and fourth transmittance dips originated from the second, third, first, and fourth Gr layers. Scattering parameters of the quadband plasmonic modulator exhibited four dips at resonant frequencies, which are provided in \textcolor{blue}{Fig. S3(a)} of \textcolor{blue}{Supplementary Material}.    
\subsubsection{Pentaband filter}
\begin{figure}[H] 
\centering\includegraphics[width=8.87cm]{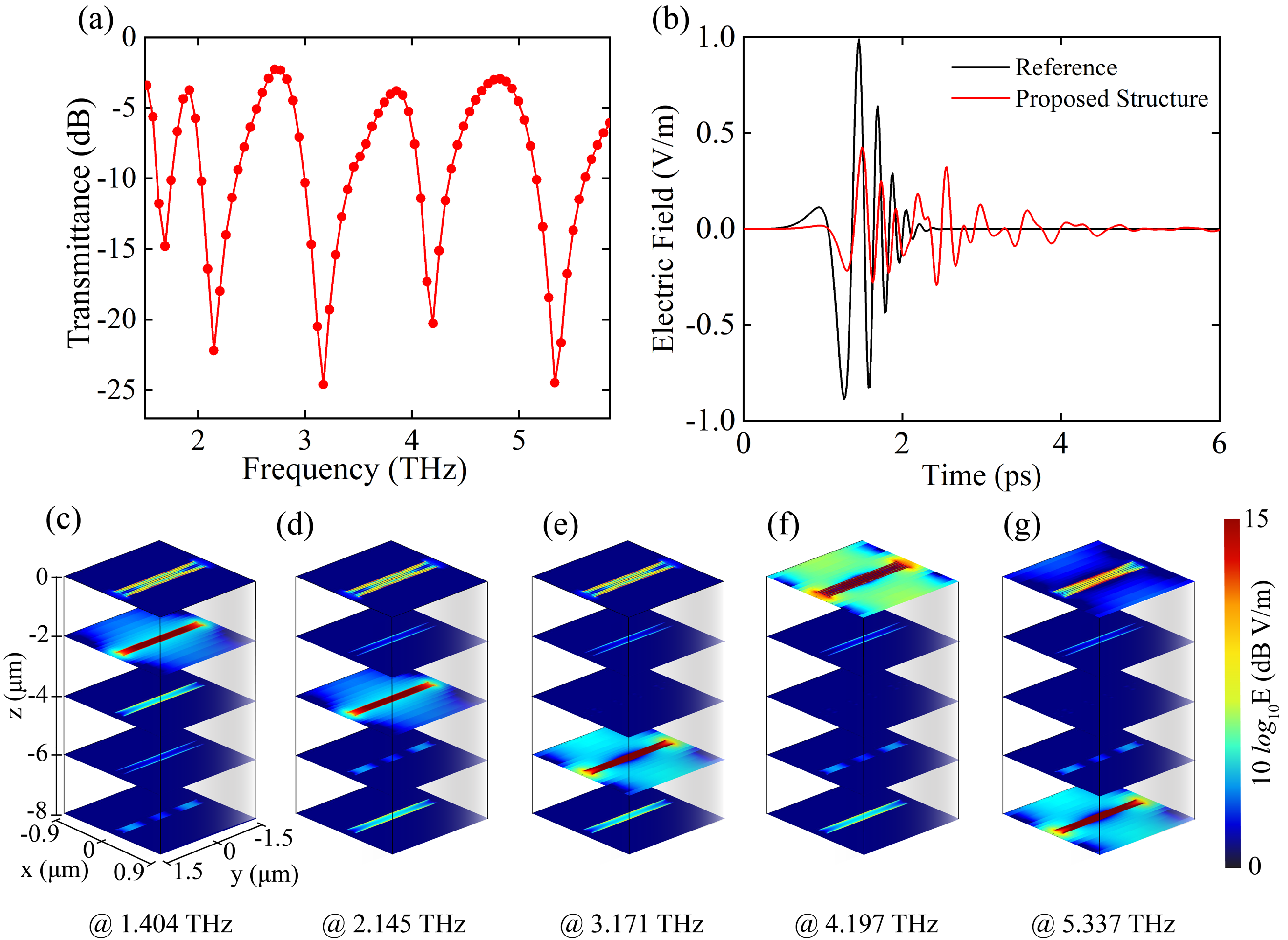}
\caption{(a) Transmittance spectra of the proposed pentaband plasmonic filter. Here, $\mu_{c_1}$ = 0.28, $\mu_{c_2}$ = 0.18, $\mu_{c_3}$ = 0.45, $\mu_{c_4}$ = 0.65, and $\mu_{c_5}$ = 0.50 eV were adopted. (b) The E-field waveforms of the proposed pentaband plasmonic filter in the time domain for TM polarized incident THz light. The spatial E-field distributions of Gr layers at resonant frequencies of (c) 1.404, (d) 2.145, (e) 3.171, (f) 4.197, and (g) 5.337 THz for our proposed pentaband plasmonic filter. The top, second, third, fourth, and bottom Gr layers exhibited localized E-fields at the resonant frequencies of 4.197, 1.404, 2.145, 3.171, and 5.337 THz, respectively. The logarithmic scale was utilized to represent the E-field color map.}
\label{Penta_filter}
\end{figure}

A pentaband plasmonic filter was modeled utilizing multilayered Gr layers with different chemical potentials. Five engineered Gr layers with different chemical potentials were employed to block light by enhancing surface light-matter interactions, as seen in Fig.~\ref{fig:1}(c). The chemical potentials of five Gr layers were set to be $\mu_{c_1}$ = 0.28, $\mu_{c_2}$ = 0.18, $\mu_{c_3}$ = 0.45, $\mu_{c_4}$ = 0.65, and $\mu_{c_5}$ = 0.50 eV. Five transmittance dips emerged at the resonant frequencies of 1.404, 2.145, 3.171, 4.197, and 5.337 THz, as shown in Fig.~\ref{Penta_filter}. Pentaband filter structure experienced a larger attenuation with prolonged time than quadband filter structure due to scattering among Gr layers and localized light field. The transmittance, reflectance, and absorptance spectra are provided in \textcolor{blue}{Fig. S2(b)} of \textcolor{blue}{Supplementary Material}. We studied the spatial electric field distributions of our proposed structure to elevate our knowledge of the localized electric fields which were depicted at resonant frequencies in Figs.~\ref{Penta_filter}(c)--(g). Scattering parameters of the pentaband plasmonic modulator exhibited five dips at resonant frequencies, as shown in \textcolor{blue}{Fig. S3(b) of Supplementary Material}. The electric field of incoming light was localized at resonant frequencies by considering each Gr layer as an attenuating medium. First, second, third, fourth, and fifth transmittance dips emerged from the second, third, fourth, first, and fifth Gr layers. 
\subsubsection{Plasmon tuning}
\begin{figure}[ht] 
\centering\includegraphics[width=8.87cm]{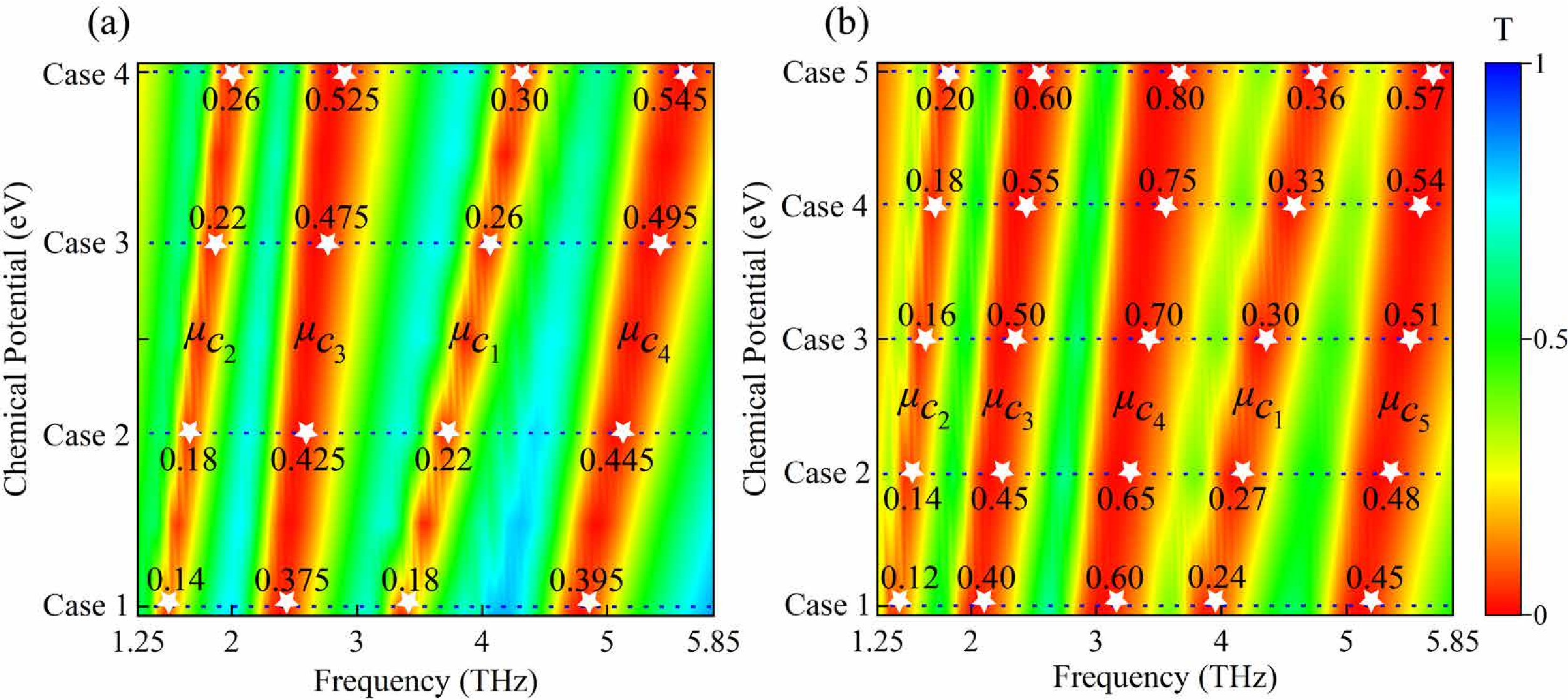}
\caption{Evolution of plasmonic (a) quadband, and (b) pentaband transmittance dips at different chemical potentials of Gr layers. We applied a TM-polarized normal incidence plane wave THz light on our proposed modulators. Transmittance dips and peaks were blue-shifted by increasing the chemical potentials of Gr layers. Dot lines represent the different case studies of our proposed modulators.}
\label{tunning_color}
\end{figure}

High tunability without structural modifications for our proposed plasmonic modulators can be achieved by changing surface conductivity via applying $V_{g_n}$. Meanwhile, the surface conductivity varied the $\mu_c$ of Gr layers~\cite{Fallah2019, Sarker:21, Sarker2023}. The relationship between $V_{g_n}$ and $\Delta \mu_c$ is given by~\cite{Gholizadeh2023},
\begin{equation}
    \Delta \mu_c\approx  |\Delta E_F| = \hbar V_F \sqrt{\frac{\pi}{e} \frac{\epsilon_0 \epsilon_{\mbox{\tiny {PMMA}}}}{t} |V_{g_n} - V_{Dirac}|}.
    \label{eq:8}
\end{equation}
Here, $\epsilon_0$, e, $\epsilon_{\mbox{\tiny {PMMA}}}$, $\Delta E_F$, and $V_{Dirac}$ represent the permittivity of vacuum, the charge of an electron, PMMA layer's permittivity, change of Fermi level, and the Dirac voltage of Gr layer, respectively. Voltage source varied the change of the Gr layer's chemical potential, as can be seen in \textcolor{blue}{Fig. S4} of \textcolor{blue}{Supplementary Material}. Here, one assumption was made that the impedance of metal contacts is negligible; therefore, it is not considered in our study. A color map of transmittance for our proposed plasmonic modulators displays the structure's wide tunability evolution at different chemical potentials, shown in Fig.~\ref{tunning_color}. The third and the fourth resonant modes of our proposed quadband and pentaband modulators, respectively, were highly tunable which were consistent with previous report~\cite{Sarker2023}. These plasmonic modulators can be operated in the wide range from 1.25 to 5.85 THz by varying the $\mu_c$ of Gr layers from 0.12 to 0.80 eV. It can be inferred from Fig.~\ref{tunning_color} that the dips in transmittance are blue-shifted when chemical potentials rise.  
\subsubsection{Impact of polarization angles}
\begin{figure}[ht] 
\centering\includegraphics[width=8.87cm]{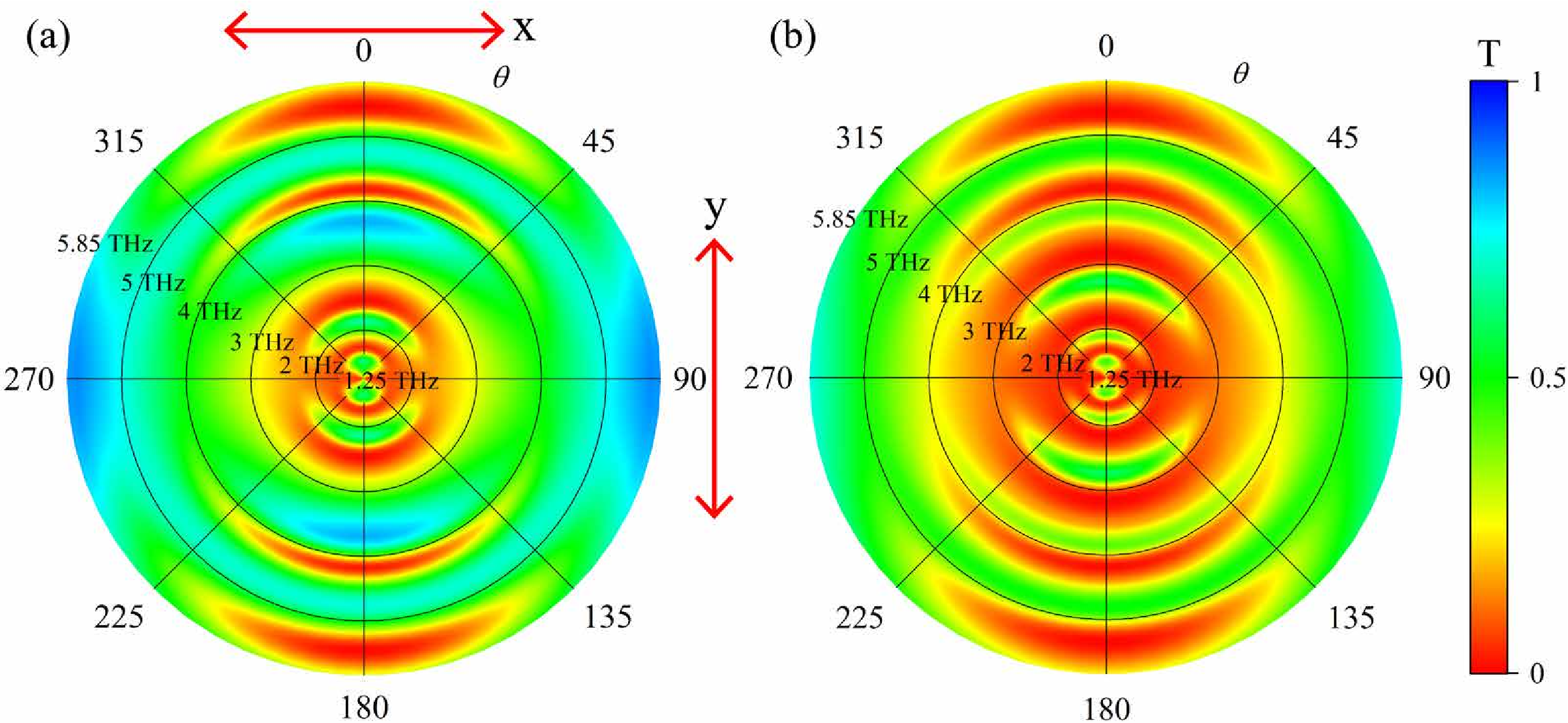}
\caption{Color plot of transmittance spectra at different $\theta$s of incoming light for our proposed (a) quadband and (b) pentaband plasmonic modulators. $\theta$ = 0$^{\circ}$ and $\theta$ = 90$^{\circ}$ represented the TM and TE polarized incident light, respectively.}
\label{angles}
\end{figure}

We analyzed the effect of the polarization angles of the incident light on the transmittance spectra of our proposed modulators, as illustrated in Fig.~\ref{angles}. Here, $\theta$ represents the polarization angle of incoming light. The electric field of TM polarized incident light ($\theta$ = 0$^{\circ}$) was strongly coupled with the proposed modulators that exhibited transmittance minima at resonant frequencies. Transmittance gradually elevated as $\theta$ increased and reached a maximum at $\theta$ = 90$^\circ$ shown in Fig.~\ref{angles}. The symmetric nature of the proposed modulators provided the same transmittance spectra at $\theta$ = 0$^\circ$ = 180$^\circ$ and $\theta$ = 90$^\circ$ = 270$^\circ$. Lower-frequency resonances were less sensitive to polarization angles compared to high-frequency resonances. It can be observed that high-frequency resonances were confined to the polarization angle range from 45$^\circ$ to 315$^\circ$ and 135$^\circ$ to 225$^\circ$. Two high-frequency resonances existed around 4.197, 5.451 THz, and 4.197, 5.337 THz for our proposed quadband and pentaband plasmonic modulators.   
\subsubsection{Performance analysis}
We enumerated the quality factor (Q) for our proposed quad and penta bandstop plasmonic structures. Q provides the relation between full-width half-maximum (FWHM) and resonant frequency, $f_c$ that is given by,
\begin{equation}
    \mbox{Q} = \frac{f_c}{\mbox{FWHM}}.
\end{equation}
Calculated Q of 4.38, 4.04, 12.27, and 7.97 were obtained at resonant frequencies of 1.746, 2.487, 4.197, and 5.451 THz for the proposed quadband plasmonic modulator. Moreover, we achieved Q of 3.7, 5.38, 5.06, 9.2, and 7.8 at resonant frequencies of 1.404, 2.145, 3.171, 4.197, and 5.337 THz for the proposed pentaband plasmonic modulator. The top GNR layer provided the highest Q compared to Gr nanoholes. Q gives the idea of the filter's selectivity. As Q increases, our filter becomes more selective with decreasing FWHM. Therefore, it can be inferred that the resonance modes at the frequencies of 4.197 and 4.197 THz exhibited high selectiveness for our proposed quadband and pentaband plasmonic filters, respectively.

Moreover, we calculated the ER, a performance parameter of the filter, which was solved by,
\begin{equation}
    \mbox{ER} = 10log(\frac{T_{TE}}{T_{TM}}).
\end{equation}
Here, $T_{TM}$ and $T_{TE}$ are the transmittances of TM and TE polarized incident light, respectively. We obtained a high difference between the transmittance under TM and TE polarized light for our proposed quadband and pentaband plasmonic modulators shown in \textcolor{blue}{Fig. S5} of \textcolor{blue}{Supplementary Material}. ERs of 4.6, 12.8, 14.4, and 22.2 dB were achieved at resonant frequencies for the quadband plasmonic filter. Similarly, we found ERs of -0.4, 10.0, 16.1, 15.2, and 22.3 dB at resonant frequencies for the pentaband plasmonic filter. 

Furthermore, we calculated the attenuation of the bandstop filters, a performance parameter of the plasmonic filter, by solving $-10log_{10}(T_{TM})$. Attenuations of 14.8, 19.1, 16.6, and 22.5 dB were obtained at resonant frequencies for the quadband plasmonic filter. Similarly, we achieved attenuations of 14.8, 22.1, 24.6, 20.3, and 24.5 dB at resonant frequencies for our proposed pentaband plasmonic filter. 
\subsection{Slow light applications}
\begin{figure}[H] 
\centering\includegraphics[width=8.87cm]{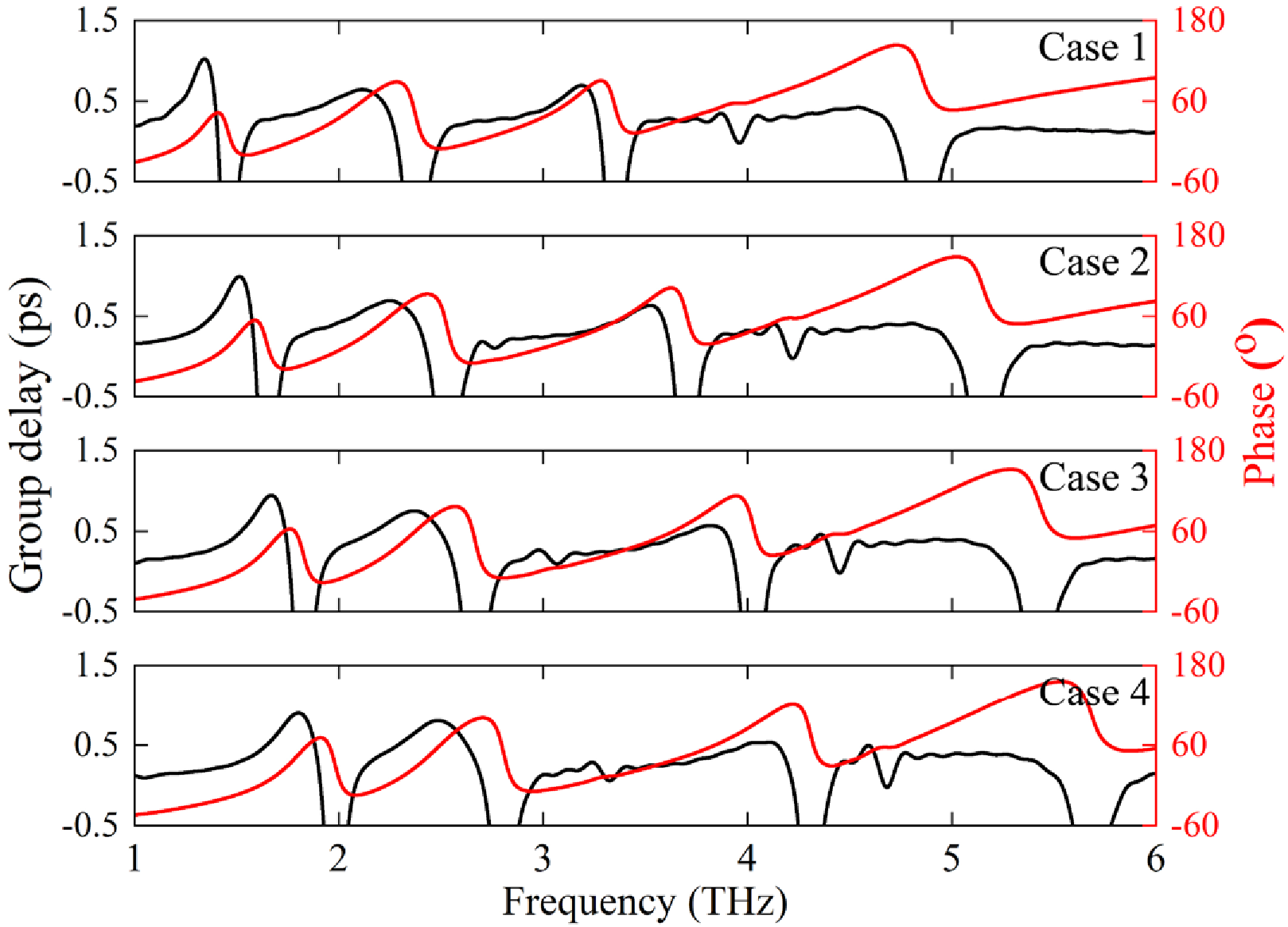}
\caption{Group delay and phase spectra of the quadband plasmonic modulator at (a) case 1, (b) case 2, (c) case 3, and (d) case 4. Black dotted lines of Fig.~\ref{tunning_color}(a) indicate the case studies of our proposed modulator.}
\label{slow_quad}
\end{figure}

\begin{figure}[ht] 
\centering\includegraphics[width=8.87cm]{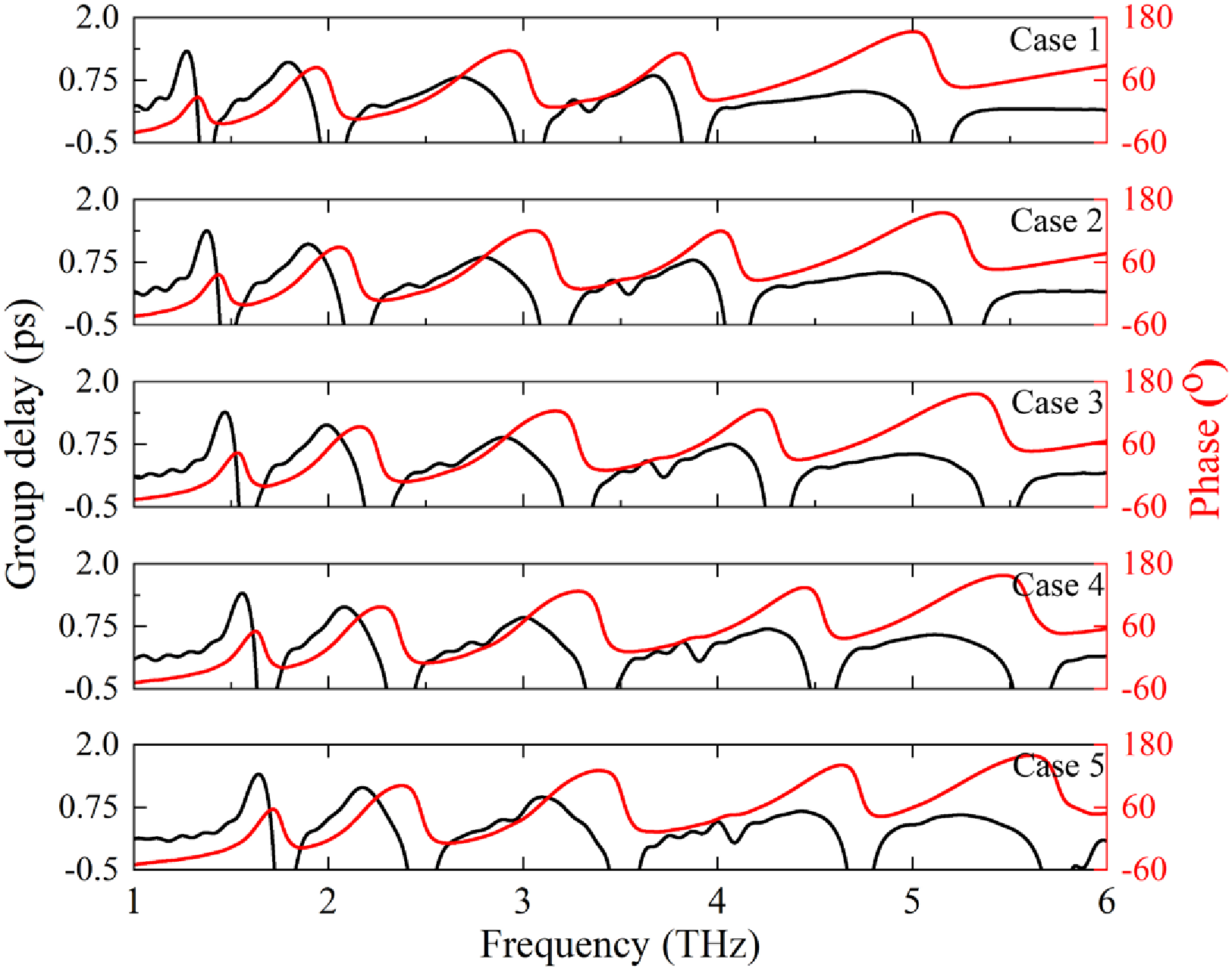}
\caption{Group delay and phase spectra of the pentaband plasmonic modulator at (a) case 1, (b) case 2, (c) case 3, (d) case 4, and (e) case 5. Dot lines of Fig.~\ref{tunning_color}(b) indicate the case studies of our proposed modulator.}
\label{slow_penta}
\end{figure}

We investigated the emergence of slow-light properties, such as group delay and phase, for our proposed THz modulator structures. Fig.~\ref{tunning_color} demonstrates the evolution of plasmonic resonances with various cases of chemical potentials at different Gr layers. Black dotted lines of Fig.~\ref{tunning_color} indicated the various cases of quadband and pentaband plasmonic modulators. The chemical potentials of different Gr layers for these cases are provided in \textcolor{blue}{Table S1} and \textcolor{blue}{S2} of \textcolor{blue}{Supplementary Material}. The transmittance peaks had a strong destructive interference effect, generating a strong dispersion with a positive group delay. Meanwhile, the phase of our proposed modulators experienced a sharp change while the frequency approached transmission peaks. This phenomenon is depicted in Figs.~\ref{slow_quad} and \ref{slow_penta}. The group delay, $\tau_g$ was calculated by,
\begin{equation}
    \tau_g = \frac{d \theta_p}{d \omega}.
\end{equation}
Here, $\omega$ and $\theta_p$ are the angular frequency of light and the phase of the transmittance coefficient, respectively. $\theta_p$ was enumerated by solving $\theta_p = arg(t)$ where $t$ denotes the modulator's transmittance coefficient. Fig.~\ref{slow_quad} illustrates the impact of changing the Gr's chemical potential on slow light properties for our proposed quadband modulator. We can change the $\tau_g$ and $\theta_p$ of our proposed quadband plasmonic modulator by varying the $\mu_{c_n}$ of Gr layers. $\tau_g$ and $\theta_p$ were blue-shifted by increasing $\mu_{c_n}$ of Gr layers, as shown in Fig.~\ref{slow_quad}. We utilized the $\mu_{c_n}$ of Gr layers from 0.12 to 0.57 eV. The chemical potentials of case 1 and case 4 were set to be $\mu_{c_1}$ = 0.16, $\mu_{c_2}$ = 0.12, $\mu_{c_3}$ = 0.35, $\mu_{c_4}$ = 0.37 and $\mu_{c_1}$ = 0.32, $\mu_{c_2}$ = 0.28, $\mu_{c_3}$ = 0.55, $\mu_{c_4}$ = 0.57 eV, respectively. The chemical potentials of the other two cases varied between the chemical potentials of case 1 and case 4. We obtained a maximum group delay of 1.02 ps for the proposed quadband plasmonic modulator.
Additionally, we calculated delay-bandwidth product (DBP) by solving DBP = $\tau_g \times \Delta f$. Here, $\Delta f$ is the bandwidth of slow light delay spectra. We require slow light devices as a buffer in integrated systems where DBP is an important performance parameter for buffer components. Slow light performance analysis of our proposed quadband plasmonic modulator is enlisted in Table~\ref{Quad_com}. We adopted cases from Fig.~\ref{tunning_color}(a) where black dotted lines indicated the four cases of our proposed quadband plasmonic modulator. DBP can be adjusted without changing structural parameters by varying $\mu_c$ via an external voltage source. A maximum DBP of 1.268 was obtained for our proposed quadband plasmonic modulator.   
\begin{table}[ht]
\centering
\caption{Calculated $\tau_g$ and DBP for various case studies of quadband plasmonic modulator}
\begin{tabular}{ 
>{\centering\arraybackslash} m{1cm}
>{\centering\arraybackslash} m{2.5cm}
>{\centering\arraybackslash} m{1cm}
>{\centering\arraybackslash} m{1.25cm}
>{\centering\arraybackslash} m{1cm}
}
        \hline
         Study & Resonant frequency (THz) & $\tau_g$ (ps) & $\Delta$f (THz) & DBP\\
        \hline
        Case 1 & 1.46 & 1.02 & 0.91 & 0.93 \\
        Case 2 & 1.64 & 0.99 & 1.09 & 1.08 \\
        Case 3 & 1.82 & 0.94 & 1.26 & 1.18 \\
        Case 4 & 1.98 & 0.90 & 1.41 & 1.27 \\
        \hline
    \end{tabular}
\label{Quad_com}
\end{table}
The impact of varying the chemical potential of Gr layers on slow light properties for our proposed pentaband plasmonic modulator was depicted in Fig.~\ref{slow_penta}. Here, we varied the $\mu_{c_n}$ of Gr layers from 0.12 to 0.80 eV in our study. Increasing $\mu_c$ of Gr layers provided blue-shifted $\tau_g$ and $\theta_p$, as can be seen in Fig.~\ref{slow_penta}. The chemical potentials of case 1 and case 5 were considered to be $\mu_{c_1}$ = 0.24, $\mu_{c_2}$ = 0.12, $\mu_{c_3}$ = 0.40, $\mu_{c_4}$ = 0.60, $\mu_{c_5}$ = 0.45 and $\mu_{c_1}$ = 0.36, $\mu_{c_2}$ = 0.20, $\mu_{c_3}$ = 0.60, $\mu_{c_4}$ = 0.80, $\mu_{c_5}$ = 0.57 eV, respectively. The $\mu_{c_n}$ of the other three cases were adopted between the $\mu_{c_n}$ of these two cases. We achieved a maximum group delay of 1.41 ps for the proposed pentaband plasmonic modulator, which was the Gr modulator's highest reported group delay. The group delay and phase were blue-shifted by increasing the Gr layers' $\mu_{c_n}$. Thus, we can alter the slow-light properties of our proposed modulators by varying the $\mu_{c_n}$ of Gr layers, which allows us to create external voltage-controlled tunable slow-light devices. 
Table~\ref{Penta_com} provides the slow light performance analysis of our proposed pentaband plasmonic modulator. We adopted cases from Fig.~\ref{tunning_color}(b) where dot lines indicated the five cases of our proposed pentaband plasmonic modulator. A calculated maximum DBP of 1.725 was achieved for our proposed pentaband plasmonic modulator.
\begin{table}[ht]
\centering
\caption{Calculated $\tau_g$ and DBP for various case studies of pentaband plasmonic modulator}
\begin{tabular}{ 
>{\centering\arraybackslash} m{1cm}
>{\centering\arraybackslash} m{2.5cm}
>{\centering\arraybackslash} m{1cm}
>{\centering\arraybackslash} m{1.25cm}
>{\centering\arraybackslash} m{1cm}
}
        \hline
         Study & Resonant frequency (THz) & $\tau_g$ (ps) & $\Delta$f (THz) & DBP\\
        \hline
        Case 1 & 1.37 & 1.329 & 0.83 & 1.103 \\
        Case 2 & 1.48 & 1.377 & 0.94 & 1.294 \\
        Case 3 & 1.57 & 1.392 & 1.03 & 1.434 \\
        Case 4 & 1.67 & 1.413 & 1.13 & 1.597 \\
        Case 5 & 1.76 & 1.414 & 1.22 & 1.725 \\
        \hline
    \end{tabular}
\label{Penta_com}
\end{table}
\section{Comparative analysis}
Table~\ref{Tab:1} shows a detailed comparative analysis of the performance parameters among the previously reported studies and our proposed modulators. Our obtained $\tau_g$ outperformed previously reported works. Xiong \textit{et al.} reported a continuous staircase-shaped structure on the Gr monolayer~\cite{D0CP06182D}. Their proposed structure provided a high ER of 16.56 dB with a $\tau_g$ of 0.488 ps. However, a staircase shape in the Gr layer makes the fabrication procedure complex. Gao \textit{et al.} proposed a modulator comprising four GNRs on silicon substrate~\cite{Gao:20}. Low ER was the limiting factor of their modulator structure. Liu \textit{et al.} demonstrated a similar kind of GNR modulator which had similar disadvantages as previous work including low ER and small operating range~\cite{Liu2020}. Hu \textit{et al.} reported an asymmetrical Gr-based structure; however, the operating range of the structure covered a small region of the THz spectrum. In comparison to existing Gr-based devices, our suggested quadband and pentaband plasmonic modulator structures offer higher ER with a $\tau_g$ of 1.41 ps which is quite large.
\begin{table}[ht]
\centering
\caption{Comparative performance analysis of Gr structures}
\begin{tabular}{ 
>{\centering\arraybackslash} m{2.75cm}
>{\centering\arraybackslash} m{1.45cm}
>{\centering\arraybackslash} m{0.98cm}
>{\centering\arraybackslash} m{0.9cm}
>{\centering\arraybackslash} m{0.9cm}
}
        \hline
         Structure & Operating range (THz) & ER (dB) & $\tau_g$ (ps) & Ref.\\
        \hline

         Multilayer graphene metamaterial & 3.91 -- 7.84 & 7.67 & 0.16 & ~\cite{Zhang:21}\\
        \hline
         Continuous patterned monolayer graphene & 2 -- 7 & 16.56 & 0.488 & ~\cite{D0CP06182D} \\
        \hline
         Dual-frequency on–off modulator & 2 -- 7 & 11.55 & 0.239 & ~\cite{Li_2020} \\
        \hline
         Patterned monolayer graphene metamaterial & 4 -- 7 & -- & 0.255 & ~\cite{Zhang:19} \\
        \hline
         Single-layer graphene ribbon and strips & 2.5 -- 5.75 & 11.75 & 0.623 & ~\cite{Gao:20} \\
        \hline
         Rectangular defect in graphene & 2.5 -- 5.5 & -- & 0.363 & ~\cite{Xu_2019}\\
        \hline
         Dual-Mode On-to-Off Modulation & 5 -- 7 & 12.55 & 0.7 & ~\cite{Liu2020} \\
        \hline
         Graphene ribbon-based grating & 4.25 -- 7.25 & -- & 0.257 & ~\cite{Xu_2021} \\
        \hline
         H-type-graphene metamaterial & 1.8 -- 7.6 & -- & 0.198 & ~\cite{Gao:19} \\
        \hline
         Asymmetrical crisscross design & 3 -- 6.5 & -- & 0.667 & ~\cite{C8CP04484H} \\
        \hline
         Split-ring resonator & 3 -- 5 & -- & 0.32 & ~\cite{Wang:22} \\
        \hline
         Multi-stacked graphene metamaterial & 1.25 -- 5.85 & 22.3 & 1.414 & This work \\
        \hline
    \end{tabular}
\label{Tab:1}
\end{table}
\section{Conclusion}
In this paper, we proposed two multilayered Gr-based quadband and pentaband plasmonic modulators and utilized the FDTD technique to conduct the performance analysis of these modulators. The Gr layers strongly interacted with the incident light's electric field, resulting in SP resonance at resonant frequencies due to scattering. Varying applied voltage changed the Gr layer's chemical potential, allowing the modulators to tune the resonant frequencies without modifying structural parameters. Meanwhile, the change in chemical potentials shifted the Gr's surface conductivity. The proposed modulators exhibited high tunability over a wide operating range of THz spectrum. We obtained a noteworthy improvement in performance for the pentaband plasmonic modulator compared to the quadband plasmonic modulator. Our designed pentaband plasmonic modulator had more than a quarter group delay compared to the quadband plasmonic modulator; however, the quadband plasmonic modulator provided a 25\% higher Q-factor. We achieved a maximum optical group delay of 1.41 ps and ER of 22.3 dB for our proposed pentaband plasmonic modulator, which is significantly higher than previously reported works. Therefore, the proposed high-performance modulators will benefit optical filtering and slow light applications, specifically in optical buffers. 
\section*{Acknowledgments}
D. S. and A. Z. thank the Dept. of EEE, Bangladesh University of Engineering and Technology (BUET), for providing technical support. D. S. acknowledges the financial support from BUET through its Postgraduate Fellowship Program. P. P. N. acknowledges the facilities from the Dept. of EEE, Shahjalal University of Science and Technology.
\bibliographystyle{IEEEtran}
\bibliography{V1}

\begin{thebibliography}{10}
\providecommand{\url}[1]{#1}
\csname url@samestyle\endcsname
\providecommand{\newblock}{\relax}
\providecommand{\bibinfo}[2]{#2}
\providecommand{\BIBentrySTDinterwordspacing}{\spaceskip=0pt\relax}
\providecommand{\BIBentryALTinterwordstretchfactor}{4}
\providecommand{\BIBentryALTinterwordspacing}{\spaceskip=\fontdimen2\font plus
\BIBentryALTinterwordstretchfactor\fontdimen3\font minus
  \fontdimen4\font\relax}
\providecommand{\BIBforeignlanguage}[2]{{%
\expandafter\ifx\csname l@#1\endcsname\relax
\typeout{** WARNING: IEEEtran.bst: No hyphenation pattern has been}%
\typeout{** loaded for the language `#1'. Using the pattern for}%
\typeout{** the default language instead.}%
\else
\language=\csname l@#1\endcsname
\fi
#2}}
\providecommand{\BIBdecl}{\relax}
\BIBdecl

\bibitem{Pendry}
J.~B. Pendry, L.~Martín-Moreno, and F.~J. Garcia-Vidal, ``Mimicking surface
  plasmons with structured surfaces,'' \emph{Science}, vol. 305, no. 5685, pp.
  847--848, 2004.

\bibitem{Gramotnev2010}
D.~K. Gramotnev and S.~I. Bozhevolnyi, ``Plasmonics beyond the diffraction
  limit,'' \emph{Nature Photonics}, vol.~4, no.~2, pp. 83--91, Feb 2010.

\bibitem{Grigorenko2012}
A.~N. Grigorenko, M.~Polini, and K.~S. Novoselov, ``Graphene plasmonics,''
  \emph{Nature Photonics}, vol.~6, no.~11, pp. 749--758, Nov 2012.

\bibitem{D}
D.~Schurig, J.~J. Mock, B.~J. Justice, S.~A. Cummer, J.~B. Pendry, A.~F. Starr,
  and D.~R. Smith, ``Metamaterial electromagnetic cloak at microwave
  frequencies,'' \emph{Science}, vol. 314, no. 5801, pp. 977--980, 2006.

\bibitem{PhysRevLett.85.3966}
J.~B. Pendry, ``Negative refraction makes a perfect lens,'' \emph{Phys. Rev.
  Lett.}, vol.~85, pp. 3966--3969, Oct 2000.

\bibitem{Sarker2023}
D.~Sarker, P.~P. Nakti, M.~I. Tahmid, M.~A.~Z. Mamun, and A.~Zubair, ``Tunable
  multistate terahertz switch based on multilayered graphene metamaterial,''
  \emph{Optical and Quantum Electronics}, vol.~55, no.~2, p. 159, Jan 2023.

\bibitem{Liu2014}
C.-H. Liu, Y.-C. Chang, T.~B. Norris, and Z.~Zhong, ``Graphene photodetectors
  with ultra-broadband and high responsivity at room temperature,''
  \emph{Nature Nanotechnology}, vol.~9, no.~4, pp. 273--278, Apr 2014.

\bibitem{Sarker:21}
D.~Sarker, P.~P. Nakti, M.~I. Tahmid, M.~A.~Z. Mamun, and A.~Zubair,
  ``Terahertz polarizer based on tunable surface plasmon in graphene
  nanoribbon,'' \emph{Opt. Express}, vol.~29, no.~26, pp. 42\,713--42\,725, Dec
  2021.

\bibitem{Li_2022}
M.~Li, Y.~Shi, X.~Liu, J.~Song, X.~Wang, and F.~Yang, ``Tunable plasmon-induced
  transparency in graphene-based plasmonic waveguide for terahertz band-stop
  filters,'' \emph{Journal of Optics}, vol.~24, no.~6, p. 065002, may 2022.

\bibitem{Xu_2020}
H.~Xu, Z.~Chen, Z.~He, G.~Nie, and D.~Li, ``Terahertz tunable optical
  dual-functional slow light reflector based on gold-graphene metamaterials,''
  \emph{New Journal of Physics}, vol.~22, no.~12, p. 123009, dec 2020.

\bibitem{Zhang2013}
L.~Zhang, J.~Yang, X.~Fu, and M.~Zhang, ``Graphene disk as an ultra compact
  ring resonator based on edge propagating plasmons,'' \emph{Applied Physics
  Letters}, vol. 103, no.~16, p. 163114, Oct 2013.

\bibitem{Zhang:21}
Z.~Zhang, Z.~Liu, F.~Zhou, J.~Wang, Y.~Wang, X.~Zhang, Y.~Qin, S.~Zhuo, X.~Luo,
  E.~Gao, and Z.~Yi, ``Broadband plasmon-induced transparency modulator in the
  terahertz band based on multilayer graphene metamaterials,'' \emph{J. Opt.
  Soc. Am. A}, vol.~38, no.~6, pp. 784--789, Jun 2021.

\bibitem{Gao:20}
E.~Gao, H.~Li, Z.~Liu, C.~Xiong, C.~Liu, B.~Ruan, M.~Li, and B.~Zhang,
  ``Terahertz multifunction switch and optical storage based on triple
  plasmon-induced transparency on a single-layer patterned graphene
  metasurface,'' \emph{Opt. Express}, vol.~28, no.~26, pp. 40\,013--40\,023,
  Dec 2020.

\bibitem{D0CP06182D}
C.~Xiong, L.~Chao, B.~Zeng, K.~Wu, M.~Li, B.~Ruan, B.~Zhang, E.~Gao, and H.~Li,
  ``Dynamically controllable multi-switch and slow light based on a
  pyramid-shaped monolayer graphene metamaterial,'' \emph{Phys. Chem. Chem.
  Phys.}, vol.~23, pp. 3949--3962, 2021.

\bibitem{Gao:19}
E.~Gao, Z.~Liu, H.~Li, H.~Xu, Z.~Zhang, X.~Zhang, X.~Luo, C.~Xiong, C.~Liu,
  B.~Zhang, and F.~Zhou, ``Dual dynamically tunable plasmon-induced
  transparency in h-type-graphene-based slow-light metamaterial,'' \emph{J.
  Opt. Soc. Am. A}, vol.~36, no.~8, pp. 1306--1311, Aug 2019.

\bibitem{Wang:22}
S.~Wang, H.~Liu, J.~Tang, M.~Chen, Y.~Zhang, J.~Xu, T.~Wang, J.~Xiong, H.~Wang,
  Y.~Cheng, S.~Qu, and L.~Yuan, ``Tunable and switchable bifunctional
  meta-surface for plasmon-induced transparency and perfect absorption,''
  \emph{Opt. Mater. Express}, vol.~12, no.~2, pp. 560--572, Feb 2022.

\bibitem{C8CP04484H}
H.~Xu, M.~Zhao, C.~Xiong, B.~Zhang, M.~Zheng, J.~Zeng, H.~Xia, and H.~Li,
  ``Dual plasmonically tunable slow light based on plasmon-induced transparency
  in planar graphene ribbon metamaterials,'' \emph{Phys. Chem. Chem. Phys.},
  vol.~20, pp. 25\,959--25\,966, 2018.

\bibitem{doi:10.1080/09500340903159495}
R.~W. Boyd, ``Slow and fast light: fundamentals and applications,''
  \emph{Journal of Modern Optics}, vol.~56, no. 18-19, pp. 1908--1915, 2009.

\bibitem{Shi:13}
Z.~Shi and R.~W. Boyd, ``Fundamental limits to slow-light
  arrayed-waveguide-grating spectrometers,'' \emph{Opt. Express}, vol.~21,
  no.~6, pp. 7793--7798, Mar 2013.

\bibitem{Zhao:19}
Z.~Zhao, H.~Zhao, R.~T. Ako, J.~Zhang, H.~Zhao, and S.~Sriram, ``Demonstration
  of group delay above 40 ps at terahertz plasmon-induced transparency
  windows,'' \emph{Opt. Express}, vol.~27, no.~19, pp. 26\,459--26\,470, Sep
  2019.

\bibitem{Tai2017}
L.~Tai, D.~Zhu, X.~Liu, T.~Yang, L.~Wang, R.~Wang, S.~Jiang, Z.~Chen, Z.~Xu,
  and X.~Li, ``Direct growth of graphene on silicon by metal-free chemical
  vapor deposition,'' \emph{Nano-Micro Letters}, vol.~10, no.~2, p.~20, Dec
  2017.

\bibitem{Abbas2014}
A.~N. Abbas, G.~Liu, B.~Liu, L.~Zhang, H.~Liu, D.~Ohlberg, W.~Wu, and C.~Zhou,
  ``Patterning, characterization, and chemical sensing applications of graphene
  nanoribbon arrays down to 5 nm using helium ion beam lithography,'' \emph{ACS
  Nano}, vol.~8, no.~2, pp. 1538--1546, Feb 2014.

\bibitem{Hanson2008}
G.~W. Hanson, ``Dyadic green's functions and guided surface waves for a surface
  conductivity model of graphene,'' \emph{Journal of Applied Physics}, vol.
  103, no.~6, p. 064302, Mar 2008.

\bibitem{Casiraghi2007}
C.~Casiraghi, A.~Hartschuh, E.~Lidorikis, H.~Qian, H.~Harutyunyan, T.~Gokus,
  K.~S. Novoselov, and A.~C. Ferrari, ``Rayleigh imaging of graphene and
  graphene layers,'' \emph{Nano Letters}, vol.~7, no.~9, pp. 2711--2717, Sep
  2007.

\bibitem{Zhuo:22}
S.~Zhuo, Z.~Liu, F.~Zhou, Y.~Qin, X.~Luo, C.~Ji, G.~Yang, R.~Yang, and Y.~Xie,
  ``Thz broadband and dual-channel perfect absorbers based on patterned
  graphene and vanadium dioxide metamaterials,'' \emph{Opt. Express}, vol.~30,
  no.~26, pp. 47\,647--47\,658, Dec 2022.

\bibitem{Lee2001}
L.-H. Lee and W.-C. Chen, ``High-refractive-index thin films prepared from
  trialkoxysilane-capped poly(methyl methacrylate)−titania materials,''
  \emph{Chemistry of Materials}, vol.~13, no.~3, pp. 1137--1142, Mar 2001.

\bibitem{Fallah2019}
S.~Fallah, K.~Rouhi, and A.~Abdolali, ``Optimized chemical potential
  graphene-based coding metasurface approach for dynamic manipulation of
  terahertz wavefront,'' \emph{Journal of Physics D: Applied Physics}, vol.~53,
  no.~8, p. 085102, Dec 2019.

\bibitem{Gholizadeh2023}
E.~Gholizadeh, B.~Jafari, and S.~Golmohammadi, ``Graphene-based optofluidic
  tweezers for refractive-index and size-based nanoparticle sorting,
  manipulation, and detection,'' \emph{Scientific Reports}, vol.~13, no.~1, p.
  1975, Feb 2023.

\bibitem{Liu2020}
Z.~Liu, E.~Gao, Z.~Zhang, H.~Li, H.~Xu, X.~Zhang, X.~Luo, and F.~Zhou,
  ``Dual-mode on-to-off modulation of plasmon-induced transparency and coupling
  effect in patterned graphene-based terahertz metasurface,'' \emph{Nanoscale
  Research Letters}, vol.~15, no.~1, p.~1, Jan 2020.

\bibitem{Li_2020}
M.~Li, H.~Li, H.~Xu, C.~Xiong, M.~Zhao, C.~Liu, B.~Ruan, B.~Zhang, and K.~Wu,
  ``Dual-frequency on–off modulation and slow light analysis based on dual
  plasmon-induced transparency in terahertz patterned graphene metamaterial,''
  \emph{New Journal of Physics}, vol.~22, no.~10, p. 103030, oct 2020.

\bibitem{Zhang:19}
B.~Zhang, H.~Li, H.~Xu, M.~Zhao, C.~Xiong, C.~Liu, and K.~Wu, ``Absorption and
  slow-light analysis based on tunable plasmon-induced transparency in
  patterned graphene metamaterial,'' \emph{Opt. Express}, vol.~27, no.~3, pp.
  3598--3608, Feb 2019.

\bibitem{Xu_2019}
H.~Xu, M.~Zhao, M.~Zheng, C.~Xiong, B.~Zhang, Y.~Peng, and H.~Li, ``Dual
  plasmon-induced transparency and slow light effect in monolayer graphene
  structure with rectangular defects,'' \emph{Journal of Physics D: Applied
  Physics}, vol.~52, no.~2, p. 025104, nov 2018.

\bibitem{Xu_2021}
H.~Xu, X.~Wang, Z.~Chen, X.~Li, L.~He, Y.~Dong, G.~Nie, and Z.~He, ``Optical
  tunable multifunctional slow light device based on double monolayer graphene
  grating-like metamaterial,'' \emph{New Journal of Physics}, vol.~23, no.~12,
  p. 123025, dec 2021.

\end{thebibliography}








\end{document}